\documentclass[preprint,graphics,dcolumn,prd,aps]{revtex4}

\begin{document}
\draft
\title{Variable G and $\Lambda$: scalar-tensor versus RG-improved cosmology}
\author{Claudio Rubano and Paolo Scudellaro}
\address{Dipartimento di Scienze Fisiche, Universit\`{a} di Napoli,\\
Complesso Universitario di Monte S. Angelo,\\ Via Cintia, Ed. N,
I-80126 Napoli, Italy\\ and \\ Istituto Nazionale di Fisica Nucleare,
Sez. Napoli,\\ Complesso Universitario di Monte S. Angelo,\\ Via
Cintia, Ed. G, I-80126 Napoli, Italy}

\begin{abstract}
We study the consequences due to time varying $G$ and $\Lambda$ in
scalar-tensor theories of gravity for cosmology, inspired by the
modifications introduced by the Renormalization Group (RG) equations
in the Quantum Einstein Gravity. We assume a power-law scale factor in
presence contemporarily of both the scalar field and the matter
components of the cosmic fluid, and analyze a special case and its
generalization, also showing the possibility of a phantom cosmology.
In both such situations we find a negative kinetic term for the scalar
field $Q$ and, possibly, an equation-of-state parameter $w_Q<-1$. A
violation of dominant energy condition (DEC) for $Q$ is also possible
in both of them; but, while in the first special case the $Q$-energy
density then remains positive, in the second one we find it negative.
\end{abstract}
\maketitle

PACS number(s): 98.80.Jk, 98.80.Cq, 98.80.Hw, 04.20.Jb

\


\newpage
\narrowtext

\section{Introduction}

In cosmology the Newton coupling term $G$ is usually assumed as
constant, as well as the cosmological term $\Lambda$. As a matter of
fact, it can be easily shown that, once we suppose a time varying
$\Lambda $, if we do not want to lose covariance of the equations we
have also to implement the time dependence of $G$. But there are also
measurements which now offer the possibility of not excluding some
variability of such important parameters; the experimental range of
variation of $G$ seems in fact to be
$10^{-12}yr^{-1}<\dot{G}/G<10^{-10}yr^{-1}$. (For some recent
considerations on this, based on white dwarf asteroseismology, see
Ref. \cite{bgi}.) It therefore seems reasonable to take
\begin{equation}
\frac{\dot{G}}{G}=\sigma H_0\,,  \label{gdot}
\end{equation}
$H_0$ being the Hubble term and $\sigma \sim 1$ \cite{1}. (As a matter
of fact, many current estimates give $\left| \sigma \right|\leq 0.1$,
but they are usually made on solar system (or at most galactic) scale.
As we are concerned here on cosmological scales, we assume a more
conservative position.) This kind of behavior is in fact obtained in
various nonstandard descriptions, for example, in scalar-tensor
theories of gravity. On the other hand, $\Lambda$ has widely been
inserted into equations as a time varying quantity (see Ref. \cite{2}
and references therein, for instance), often without any consideration
on covariance and the contemporary variability of the $G$-term.
Anyway, we can suppose \cite{1,2,23,24} that $\Lambda_0\sim{H_0}^{2}$,
also in order to explain current astrophysical observations, which
lead, as known, to dark energy models of the universe.

Also, due to recent work on cosmology in the Planck era, soon after
the initial Big Bang singularity, some models have been discussed
\cite{3}, where the Newton constant $G$ and the cosmological constant
$\Lambda$ may be dynamically coupled to the geometry of spacetime. For
instance, the Einstein equations can be modified by the
renormalization group (RG) equations for Quantum Einstein Gravity
\cite{4}, solving horizon and flatness problems of the cosmological
standard model, without need of introducing inflation. The use of a
dark energy scalar field to solve the {\em cosmic coincidence} problem
is also not needed when the large scale dynamics of the universe is
studied in such a framework, since the vacuum energy density,
depending on $G$, can be automatically equalled to the matter energy
density, so that $\Omega_{\Lambda}=\Omega_{m}=1/2$ \cite{5}. In such a
context it results $\sigma\equiv(3/2)(1+w_{m})$, $w_{m}$ being the
ratio of the pressure to the energy density of the matter component
\cite{prep}.

Different approaches are possible. For example, one can try to set the
improvement at the level of the action \cite{reuterw}. (See comments
about this approach in the concluding remarks.) It is also possible to
treat $G$ and $\Lambda$ as dynamical variables in the action from the
very beginning \cite{adm}.

Here, we want to show that some of what is gained in an approach like
that in Refs. \cite{3,4,5} can also be obtained in a scalar-tensor
theory of gravity, and comment on this. Let us first stress that in
such a kind of models both a time-varying effective $G$ and a {\em
natural} effective cosmological $\Lambda$-term, also varying with
time, may be introduced \cite{6,7,8}. Let us thus consider a
nonminimally coupled theory for cosmology, where a scalar field $Q$
couples to geometry via a function $F=F(Q)$; in the energy content of
the universe we also include nonrelativistic matter, which is supposed
to evolve substantially decoupled from $Q$. This kind of model gives
the possibility to describe the cosmological situation at least
starting from the decoupling period. As a matter of fact, therefore,
we expect to recover some of the features characterizing the nowadays
debate on present cosmology. In this paper we do not aim at developing
our approach completely, but we anyway think we can at least grasp the
most relevant aspects (like, for instance, the present cosmic
acceleration) as dynamical results of the model. As a matter of fact,
we also find the possibility to recover a phantom cosmology, with a
negative kinetic term but positive definite energy density, and an
equation of state $w_Q<-1$.

Phantom matter is usually defined as that with equation-of-state
parameter $w_Q<-1$ \cite{81}. It was introduced in the context of
cosmology assuming a minimal coupling between scalar and gravitational
fields, when such limits on $w_Q$ are directly implied by taking
$\rho_Q>0$, $V(Q)>0$, and ${\dot{Q}}^2<0$. It always results $p_Q<0$,
too. From this all, the dominant energy condition (DEC)
$\rho_Q+p_Q\geq0$ \cite{82} results also violated \cite{81,83},
leading to a situation that has been not often considered in standard
cosmology, with some difficulties in making the phantom model stable
\cite{83}. (See Ref. \cite{84} for a possible finite-time DEC
violation, arising from a bulk viscous stress due to a particle
production.) It is also interesting to note that scalar fields with
negative kinetic energies \cite{81,83} were not used in minimally
coupled cosmology since the old Steady State theory
\cite{85,86,87,88}, where a creation field with such a feature was in
fact introduced \cite{89}. (For some comments on $\rho_Q<0$, see Refs.
\cite{89,salgado,french}, for instance.)

Nonetheless, it has to be stressed that the observational limits on
$w_Q$ give $-1.62<w_Q<-0.74$ at the $95\%$ confidence level \cite{810}
or $-1.18<w_Q<-0.93$, using also the Wilkinson Microwave Anisotropy
Probe (WMAP) data \cite{811}, so allowing the possibility of what is
now called {\em superquintessence} \cite{812} or {\em phantom energy}
\cite{81}. There is also evidence for some consistency between the
current cosmic age estimator data and phantom energies \cite{811,817}.
A discussion on whether this component of the universe may lead to a
{\em big rip} \cite{813,814} or not \cite{815,816} is still open, even
if the occurrence of a possible singularity in the next future has
recently been discussed also when $\rho_Q>0$ and $\rho_Q + 3p_Q>0$
\cite{817bis}.

On the other hand, when there is a nonminimal coupling between $Q$ and
geometry, we can have negative kinetic energies for $Q$ without having
$w_Q<-1$ or $\rho_Q + p_Q<0$, even assuming $\rho_Q>0$ and $V(Q)>0$.
This should be connected with the fact that now $\rho_Q$ and $p_Q$
have more complicated expressions, strongly depending on the time
variation of the coupling function $F(Q(t))$ \cite{8}. As a matter of
fact, phantom energy has been already considered in the context of
scalar-tensor theories \cite{814,818,819}. (In Ref. \cite{817} it is
shown that a time dependent $w_Q$ is more favourable to the existence
of epochs with $w_Q<-1$, while in Ref. \cite{814} it is also shown
that a big rip in the future evolution of the universe can be avoided
for some range of the parameters involved once $w_Q=w_Q(t)$ is
assumed, and it probably results unavoidable with a constant $w_Q$
parameter.)

Here, our aim is to assume $G$ and $\Lambda$ variations inspired by the RG
improved gravity and comment on results deriving from plugging them into the
general framework of a scalar-tensor theory with Lagrangian density
\begin{equation}
{\cal {L}}=\sqrt{-g}\left[
F(Q)R+\frac{1}{2}g^{\mu\nu}Q_{;\mu}Q_{;\nu}-V(Q)
\right]+{\cal {L}}_m\,,
\end{equation}
where $F(Q)$ and $V(Q)$ are the generic functions respectively describing
the coupling and the potential of the scalar field $Q$; $R$ is the scalar
curvature and the metric is the Friedmann-Lemaitre-Robertson-Walker (FLRW)
one. We represent with ${\cal {L}}_m$ the usual matter contribution, seen as
decoupled from the scalar sector, and the standard minimally coupled
situation is recovered when $F=-1/2$, with units $8{\pi}G=c=1$.

The use of power-law solutions for the scale factor $a=a(t)$ makes it clear
that it is possible to have a phantom behavior only for special choices of
the parameters involved. This has already been shown elsewhere \cite{814}.
But we can here show that we cannot simply neglect ordinary matter. Even if
a {\em superacceleration} $\ddot{a}>0$ implied by an equation-of-state
parameter $w_Q<-1$ is supposed to lead to a regime where the scalar field $Q$
dominates \cite{811,814}, it in fact appears interesting to better
investigate the situation, since the range of parameters allowing a phantom
cosmology results tightly connected with the present dust content of the
universe.

In the following, in Sec. II we give the basics of the nonminimally
coupled theory used in our discussion, while in Sec. III we
investigate a special (and motivated) choice of $H=H(t)$. Through a
suitable generalization of some parameters, in Sec. IV we try to
discover new and more general features of the model. Since we know
that there must be a connection between our {\em Jordan-frame}
formulation in Sec. III and the usual minimally coupled equations of
cosmology in presence of a scalar field (i.e., the {\em
Einstein-frame} formulation), Sec. V is dedicated to investigate the
expression of the particular conformal transformation that realizes
this. In Sec. VI, in the end, we trace the final comments and
conclusions.

\section{Nonminimally coupled theory}

In presence of a nonminimally coupled scalar field $Q$ in the
universe, as already said, we have to consider an effective
gravitational coupling, taking into account the fact that, now,
geometry (and, thus, the gravitational field) is sensibly coupled to
this new component. In what follows we only take two components
filling the universe, i.e., matter and scalar field. By ``matter'' we
here mean ``dust'' (cold noninteractive matter, i.e., baryons and dark
matter together), so that the period involved by our description is
placed substantially after the decoupling era, when the radiation
component decreases much more than the matter one. On its side, the
scalar field behavior is ruled by a potential $V(Q)$, which heavily
enters into the equations, complicating their solutions. As to that,
in what follows we only limit ourselves to write down the equations
needed for our discussion. Further details can be found, for example,
in Ref. \cite{8}.

Let us start from the Einstein and Klein-Gordon equations
\begin{equation}
3H^{2}={\cal G}\left( \rho_{m}+\rho_{Q}\right)\,,  \label{1}
\end{equation}
\begin{equation}
\ddot{Q}+3H\dot{Q}+12H^{2}F^{\prime}+6\dot{H}F^{\prime}+V^{\prime}=0\,,
\label{2}
\end{equation}
where dot stands for time derivative, the prime indicates derivative with
respect to $Q$, the function $H=H(t)$ is the Hubble parameter, $\rho_{m}$ is
the matter energy density, and
\begin{equation}
\rho_{Q}=\frac{1}{2}\dot{Q^{2}}+V+6H\dot{F}\,;  \label{3}
\end{equation}
we can also define
\begin{equation}
p_{Q}=\frac{1}{2}\dot{Q}^{2}-V-2\left( \ddot{F}+2H\dot{F} \right)\,.
\label{4}
\end{equation}
In these equations we have set as usual $8\pi G = c = 1$ and introduced an
effective gravitational coupling ${\cal G}$ such that
\begin{equation}
{\cal G}=-(2F)^{-1}\,,  \label{4bis}
\end{equation}
$F=F(Q)$ being the coupling parameter of the theory; it can be shown that an
effective $\Lambda$-term can also be introduced \cite{8} such that
\begin{equation}
\Lambda_{eff}={\cal G}\rho_{Q}\,.  \label{5}
\end{equation}
This, of course, implies that Eqs. (\ref{1}) and (\ref{2}) become
\begin{equation}
3H^{2}={\cal G}\rho_{m}+\Lambda_{eff}\,,  \label{6}
\end{equation}
\begin{equation}
\dot{Q}(\ddot{Q}+3H\dot{Q}+V^{\prime})+3\frac{{\cal G}^{\prime}}{{\cal G}^2}
(2H^2+\dot{H})=0\,.  \label{6bis}
\end{equation}

We can easily see that
\begin{equation}
\dot{\rho}_{Q}+3H\left( p_{Q}+\rho_{Q} \right) = -6H^{2}\dot{F}  \label{7}
\end{equation}
and
\begin{equation}
\dot{\Lambda}_{eff}+\dot{{\cal G}}\rho_{m}=-3HG\left(
p_{Q}+\rho_{Q}\right)\,.  \label{8}
\end{equation}
This last equation has to be compared with the additional equation of the RG
improved cosmology
\begin{equation}
\dot{\Lambda}_{eff}+\dot{{\cal G}}\rho_{m}=0\,.  \label{9}
\end{equation}
We see that, if we want to interpret $\Lambda_{eff}$ {\em strongly}, as a
varying cosmological term, we have to make the additional assumption
\begin{equation}
p_Q=-\rho_Q\,,  \label{9bis}
\end{equation}
which still fits the dominant energy condition $p_Q+\rho_Q\geq0$ on the
scalar field sector. This point marks a relevant difference with the minimal
coupling case. There, in fact, Eq. (\ref{9bis}) just leads to a constant
potential, i.e., a genuinely constant $\Lambda$. In our case, the situation
is clearly different, and from Eqs. (\ref{3}), (\ref{4}), and (\ref{9bis})
it is easy to derive
\begin{equation}
(\dot{Q})^2=2(\ddot{F}-H\dot{F})\,.  \label{9ter}
\end{equation}
In the following, therefore, we should distinguish between
$\Lambda_{eff}^{weak}$ and $\Lambda_{eff}^{strong}$, as we may name
the different expressions taken by the $\Lambda$-term in the two
possible situations.

On the other hand, another important Einstein equation holds
\begin{equation}
2\dot{H}+3H^{2}=-{\cal G}\left( p_{m}+p_{Q}\right) \,,  \label{10}
\end{equation}
which implies that
\begin{equation}
\ddot{a}>0 \Longleftrightarrow \left( \rho_{m}+3p_{m} \right) +\left(
\rho_{Q}+3p_{Q} \right) <0\,.  \label{11}
\end{equation}
Since we introduce the equations of state
\begin{equation}
p_{m}=w_{m}\rho_{m},\quad p_{Q}=w_{Q}\rho_{Q}\,,  \label{12}
\end{equation}
with $w_m = 0$, it is then customary to deduce that the universe lives in an
acceleration stage when
\begin{equation}
w_{Q}<-\frac{1}{3}\,.  \label{13}
\end{equation}
As it seems generally accepted, observational results accounting for
the most recent history of the universe indicate that this is just the
case \cite{9,9a,9b} and that right now, therefore, we are in a sort of
{\em soft} inflationary epoch.

\section{A special case}

The equations introduced above are hard to deal with, but can anyway
be solved exactly in some cases \cite{8}. For our purposes we here
limit ourselves to postulate a given special solution (certainly
leading to acceleration) and speculate on consequences. To choose a
suitable one, even in a scalar-tensor theory, we again refer to RG
improved cosmology \cite{5}, with which some of our considerations
have to be naturally compared, and especially to the fact that the
infrared (IR) fixed point model for cosmology fits the high redshift
observations on Type Ia Supernovae \cite{9bis}.

Thus, we investigate the possibilities involved by setting
\begin{equation}
H=\frac{4}{3}t^{-1}\,,  \label{14}
\end{equation}
together with
\begin{equation}
{\cal G}=\alpha t^{2}  \label{15}
\end{equation}
($\alpha$ being an unknown parameter). Referring to Eq. (\ref{gdot}) we see
that this sets $\sigma=3/2$. (As said, this rather high value cannot be
excluded by present observations. See also Ref. \cite{prep} for a deeper
discussion on this point.)

Due to Eq. (\ref{5}), we can also easily see that it fits the behavior
there recalled for $\Lambda$. As a matter of fact, Eqs. (\ref{4bis})
and (\ref{14}) give
\begin{equation}
F=-\frac{1}{2\alpha }t^{-2}\,,  \label{16}
\end{equation}
and
\begin{equation}
a(t)\sim t^{\frac{4}{3}}\Longrightarrow \rho_{m}\sim a^{-3}\sim t^{-4}\,,
\label{17}
\end{equation}
since we are assuming dust as matter. Let us, therefore, pose
\begin{equation}
\rho_{m}=Mt^{-4}\,,  \label{18}
\end{equation}
with $M\equiv \rho_{m0}t_{0}^{4}$ a positive constant determined by the
present values of matter energy density and time. From Eq. (\ref{1}) we
easily get
\begin{equation}
\rho_{Q}=\left( \frac{16}{3\alpha }-M\right) t^{-4}\,,  \label{19}
\end{equation}
which is positive when
\begin{equation}
M<\frac{16}{3\alpha}\,.  \label{20}
\end{equation}
So, we find that, due to Eqs. (\ref{18}) and (\ref{19}), the two time
behaviors for $\rho_{m}$ and $\rho_{Q}$ are equally time scaling.
Allowing values of the parameters such that $M>16/(3\alpha)$ would
imply $\rho_Q<0$, leading to a situation that we choose not to discuss
here. Also, from Eq. (\ref{5}) we soon have
\begin{equation}
\Lambda_{eff} = 3\left( 1 -\frac{3M\alpha}{16} \right) H^{2}\,,
\label{lambda}
\end{equation}
that is, $\Lambda_{eff} = \lambda H^{2}$, with $0 < \lambda < 3$, which is
an acceptable order of magnitude (see Ref. \cite{1}, for example).

Substituting Eqs. (\ref{3}) and (\ref{16}) into Eq. (\ref{19}), we find an
equation which can be differentiated, so that, taking also Eq. (\ref{2})
into account, we finally get
\begin{equation}
\dot{Q}^{2}=-\left( \frac{6}{\alpha}+M\right) t^{-4}\,.  \label{21}
\end{equation}
This means that, even if the energy density associated to $Q$ is positive,
the kinetic energy term is negative and $\dot{Q}$ is imaginary; all the
other quantities used in the following, and which are more significant in
the development of our considerations, anyway result real.

On the other hand, let us note that in Ref. \cite{5} there is already the
reference to some solutions in a Brans-Dicke (BD) model \cite{12}, which
also present something similar but {\em disguised}, since in that context a
negative $\omega$ BD parameter is found to be necessary, in order to get an
accelerated phase of expansion.

Let us go on, in our own context, underlining that, due to Eq.
(\ref{3}) there is {\em no} need to have $V>0$ in order to get
$\rho_{Q}>0$. We in fact find (see Eq. (\ref{28bis2} below) that it
can be negative. From Eq. (\ref{21}) we thus set
\begin{equation}
Q=Q_{0}-i\sqrt{\frac{6}{\alpha}+M} t^{-1}\,,  \label{25}
\end{equation}
$Q_{0}$ being an integration constant that we can safely set to zero.

In a minimally coupled situation, this classical rotation of $Q$ to
imaginary values can, for example, be considered as typical of an axionic
component for vacuum phantom energy, which violates the DEC and has an
increasing energy density (so allowing the possibility of a big rip), at
least if one wants to keep the weak energy condition (WEC) \cite{13}. Here,
anyway, it is important to note that, due to Eq. (\ref{19}), the
nonminimally coupled situation implies always {\em decreasing} energy
density with time, meaning that caution is needed when simply importing
features from the minimally coupled case. (See Ref. \cite{814} for a special
nonminimally coupled model also exhibiting decreasing phantom energy.)

We can of course extract $F(Q)$ from Eq. (\ref{16})
\begin{equation}
F(Q)=\frac{1}{2(6+M\alpha)}Q^{2}< 0  \label{28}
\end{equation}
and $V(Q)$ from Eq. (\ref{3}). We have first to take into account that
\begin{equation}
V(t)=\left( \frac{1}{3\alpha}-\frac{M}{2}\right)t^{-4}\,,  \label{28bis1}
\end{equation}
which is non negative when $M\leq 2/(3\alpha)$ and negative for $
2/(3\alpha)<M<16/(3\alpha)$ (due to Eq. (\ref{20})); thus, we finally
find
\begin{equation}
V(Q)=\frac{\alpha(2-3M\alpha)}{6(6+M\alpha)^2}{Q}^4\,.  \label{28bis2}
\end{equation}
Let us stress that our initial guess in Eq. (\ref{14}) has clearly to
be meant as expressing an {\em asymptotic} regime. Thus, the above
expressions for $F$ and $V$ are strictly valid only in this
approximation. This explains why we do not recover the ordinary
general relativistic theory with $F=-1/2$. (In order to obtain this,
we should improve the approximation, which will be done in a
forthcoming paper.)

We can also get the expression of pressure from Eq. (\ref{4})
\begin{equation}
p_{Q}=-\frac{8}{3\alpha }t^{-4}\,,  \label{29}
\end{equation}
so that it always results $p_{Q}<0$. This soon gives
\begin{equation}
w_{Q}\equiv \frac{p_{Q}}{\rho _{Q}}=-\frac{8}{16-3M\alpha }=const.<0\,,
\label{30}
\end{equation}
due to Eq. (\ref{20}), which represents a necessary constraint in our
treatment. The position $-1\leq w_{Q}<0$, on the other hand, implies
$M\leq 8/(3\alpha )$. But, of course, what can be most interesting
here is the possibility that $w_{Q}<-1$, a range of values usually
involved by the negative kinetic term for the scalar field
\cite{81,13bis}. As a matter of fact, such a feature is realized when
$M>8/(3\alpha )$. Thus, due to Eqs. (\ref{28bis1}) and (\ref{28bis2}),
we find $w_{Q}<-1$ only for negative values of $V$, i.e., when
$8/(3\alpha )<M<16/(3\alpha)$.

We can schematically summarize the most important informations in
Table I. The third case there recorded (characterized by a constant
negative potential) gives a strong $\Lambda _{eff}$, according to the
above discussion, and is strictly equivalent to the RG improved
environment (see below), which is at the very origin of this
treatment. But, as should have been expected, we can see that the
situation here is richer. For example, we find a phantom energy
presence in the fourth situation. We also see that, remarkably, the
presence of matter ($M\neq 0$) is crucial in understanding what
happens. Situations differs from one another depending on suitable
ranges of the $M$ parameter, which is peculiar to characterize the
amount of ordinary matter in the cosmic content. So, we can find DEC
violation, but only for a range of $M$ values depending on the
$\alpha$-parameter characterizing the scale of time variation of the
running gravitational $G$-term, or we should rather say that such a
violation is obtained for a range of $\alpha$-values constrained by
the present-day matter content in $M$.

A final interesting insight is given by the calculation of the matter
density parameter. We find a constant matter density universe
\begin{equation}
\Omega_{m}\equiv \frac{G\rho_{m}}{3H^{2}}=\frac{3\alpha M}{16} < 1\,,
\label{32}
\end{equation}
due to Eq. (\ref{20}). When $M=8/(3\alpha)$ it is $\Omega_{m}=1/2$;
such a value is in fact produced not using the Newtonian coupling
parameter $G_N$, but through an effective scale dependent $G$, as
discussed in Ref. \cite{5}. As a matter of fact, we do find a complete
correspondence with the RG-improved solutions in Ref. \cite{5}, once
we pose $\alpha=3\pi g_{*}\lambda_{*}$. However, our case is somewhat
more general, since we also consider $w_Q \neq -1$ and always get a
negative $\dot{Q}^2$-term.

\section{A family of more general cases}

We want now to show that the ansatz in Sec. III, although inspired by a
natural guess in a very different context, reveals to have very nice
features also in this situation. In order to see this we try to release
conditions in Eqs. (\ref{14}) and (\ref{15}), and set more generally
\begin{equation}
H=nt^{-1}\,,\quad G=\alpha t^{m}\,,  \label{45}
\end{equation}
where $m>0$ and $n>1$. What we discussed in the previous section is then
relative to the special values $n=4/3$ and $m=2$. Here, we want to
generalize the behaviors analyzed before and see what happens, which
features are retained and which are not, with a position that corresponds to
releasing the constraint $\Lambda {\cal G} = const.$ of the RG improved
approach.

As a first thing, let us note that Eq. (\ref{45}) is consistent with what
one can expect for $\dot{{\cal G}}/G$ (that is, its proportionality to $H$,
through a factor $\sigma=m/n$). Later on we will anyway see that something
more careful has to be said about $\Lambda$.

From Eq. (\ref{1}) we now find
\begin{equation}
\rho_{Q}=\left( \frac{3n^{2}}{\alpha}t^{3n-m-2}-M\right) t^{-3n}\,,
\label{50}
\end{equation}
so that
\begin{equation}
\rho_{Q}>0\Longrightarrow t^{3n-m-2}>\frac{\alpha M}{3n^{2}}\,.  \label{51}
\end{equation}
When $m=3n-2$, this condition on scalar field energy density leads to
$3n^{2}>\alpha M$, i.e., to a generalization of the constraint in Eq.
(\ref{20}) found for $n=4/3$ (and $m=2$). If $m<3n-2$, it gives
\begin{equation}
t>\left( \frac{\alpha M}{3n^{2}}\right)^{\frac{1}{3n-m-2}}\,,  \label{52}
\end{equation}
so that it results $\rho_{Q}>0$ only when $t>\bar{t}\equiv \left(
\frac{\alpha M}{3n^{2}}\right)^{\frac{1}{3n-m-2}}$. If instead
$m>3n-2$, we get
\begin{equation}
t<\left( \frac{3n^{2}}{\alpha M}\right)^{\frac{1}{-3n+m+2}}\,,  \label{53}
\end{equation}
which does not allow to have $\rho_{Q}>0$ after the time $\tilde{t}\equiv
\left( \frac{3n^{2}}{\alpha M}\right)^{\frac{1}{-3n+m+2}}$.

Only the first two situations ($m \leq 3n-2$) will be considered here, since
the third one ($m>3n-2$) seems to be possibly appropriate only for an
earlier epoch (but then we should consider radiation + matter + scalar
field). Therefore, in the following we shall assume
\begin{equation}
3n\geq m+2\,,\,\,\,\, t>\tilde{t}\,.  \label{54}
\end{equation}

Performing exactly the same steps as in the previous section, we can now see
that
\begin{equation}
\dot{Q}^{2}=-\left\{ \frac{1}{\alpha}\left[ m^{2}+ (n+1)m-2n\right]
t^{3n-m-2}+M\right\} t^{-3n}\,.  \label{55}
\end{equation}
In the case when $3n=m+2$, and taking into account $n>1$ and $M>0$, it
is easy to check that $\dot{Q}^{2}<0$ again. This case is illustrated
below (see subsection A3). When $3n> m+2$, the first term dominates
more and more at increasing times. As to the sign of the coefficient
of such a term in $t^{-m-2}$, since $\alpha>0$ it is easy to see that
it is non negative when
\begin{equation}
m\geq\tilde{m}\equiv-\frac{1}{2}\left(n+1-\sqrt{(n+1)^{2}+8n}\right)\,.
\label{56}
\end{equation}
This is in agreement with Eq. (\ref{54}), implying $n>2/3$, since we
have chosen $m>0$ from the very beginning of our considerations; this
means that (being $n>1$) it has to be $m\geq\tilde{m}$ (with
$\sqrt{3}-1<\tilde{m}<2$), in order to get a negative kinetic term for
the scalar field. In particular, we may note that the special value
$m=\tilde{m}$ leads to $\dot{Q}^{2}=-Mt^{-3n}<0$, that is, to a kind
of behavior of the function $\dot{Q}^{2}$, which is clearly like the
one previously found when $n=4/3$ and $m=2$. This means that the
generalization we are examining here for $m\geq\tilde{m}$ can still
lead to a phantom cosmology. It remains to see whether the other
related features we found before are also reproduced in this more
generalized context.

As a first step, we can use Eq. (\ref{3}) to find
\begin{equation}
V\left( Q(t) \right) =\frac{1}{2}\left[ \frac{1}{\alpha }(
6n^{2}+m^{2}-5nm-2n+m) t^{3n-m-2}-M\right] t^{-3n}\,.  \label{57}
\end{equation}
Note that if we want the potential to be positive after some time $t^{*}$,
then it is necessary that
\begin{equation}
6n^{2}+m^{2}-5nm-2n+m>0\,,  \label{58}
\end{equation}
implying
\begin{equation}
m<m_{-}\equiv 2n\,,\,\, or\,\,\,\quad m>m_{+}\equiv 3n-1\,;  \label{59}
\end{equation}
thus, for $n>1$, we find
\begin{equation}
m<2n\,.  \label{60}
\end{equation}
This result is more restrictive than the one in Eq. (\ref{54}), even if
certainly does not contradict the one in Eq. (\ref{56}).

More difficult is here to get $V=V(Q)$, since it results (suitably choosing
the sign)
\begin{eqnarray}
Q(t) = -\frac{2t\sqrt{At^{-m-2}-Mt^{-3n}}}{3n-2} \,\,\,\,\,\,\,\,\,\,\,\,\,
\,\,\,\,\,\,\,\,\,\,\,\,\,\,\,\,\,\,  \nonumber \\
+\left[ \frac{2A(2+m-3n)t^{-m-1}}{(3n-2)(2+2m-3n)}\sqrt{
\frac{M-At^{3n-m-2}}{M(At^{-m-2}-Mt^{-3n})}}\,\,\Phi \right]\,,
\label{61}
\end{eqnarray}
where $A=A(n,m)$ is a constant depending on the values for $\alpha$, $n$,
and $m$
\begin{equation}
A(n,m)\equiv -\frac{1}{\alpha}(m^{2}+nm-2n+m)\,,  \label{62}
\end{equation}
and $\Phi=\Phi(n,m)$ is a hypergeometric function \cite{slater} depending on
time and all the parameters introduced
\begin{equation}
\Phi(n,m)\equiv {}_2F_1\left[ \frac{2+2m-3n}{4+2m-6n}\,,\frac{1}{2}\,; \frac{6+4m-9n}
{4+2m-6n}\,;\frac{At^{3n-m-2}}{M} \right]\,.  \label{63}
\end{equation}
This means, in fact, that it is in general impossible to find the
analytical form of the function $t=t(Q)$, that should be substituted
into Eq. (\ref{57}), so as to deduce $V=V(Q)$. However, we can easily
choose some suitable special values and see what happens at least in
the related situations.

\subsection{Some specific choices}

In the following, we will in fact deal with three subcases, since they soon
result analytically treatable and/or physically meaningful. We choose to
describe only the essentials we can easily deduce from them, being our main
purposes purely indicative here with respect to the still possible full
treatment we could perform with numerical techniques, within a context that
should certainly be much more refined.

\subsubsection{$M=0$ (i.e., complete scalar field dominance)}

When the dust component can be neglected (for example, in a late time
superaccelerating stage of complete scalar field dominance), we have
that $M=0$. So, Eq. (\ref{50}) gives an always positive scalar-field
energy density for any values of $n$ and $m$. Also, due to Eq.
(\ref{56}), we see that ${\dot{Q}}^2<0$ when $m\geq\tilde{m}$, which
implies the possibility to have parameter values leading to a phantom
behavior of the scalar field cosmology.

More in detail, with $M=0$, from Eq. (\ref{55}) it results that
${\dot{Q}}^2=At^{-m-2}$. Note also that $A<0$ when $m>\tilde{m}$. (The
situation with $m=\tilde{m}$ is equivalent to choosing $A=0$ and leads
to a constant scalar field; since we have also posed $M=0$, we will
not discuss it here.) We have $\dot{Q}={\pm}i\sqrt{|A|t^{-m-2}}$ and find
an imaginary scalar field as a function of time (not being important
the precise sign we choose). It can be soon inverted to
\begin{equation}
t(Q)=i^{2/m}\left[ \frac{2\sqrt{-A}}{m} \right]^{2/m}Q^{-2/m}\,.  \label{A2}
\end{equation}
Inserting Eq. (\ref{A2}) into $F(t)=-t^{-m}/(2\alpha)$, we can now
find $F=F(Q)$ as a quadratic function
\begin{equation}
F(Q)=-\frac{m^2}{8\alpha A}Q^2 <0  \label{A21}
\end{equation}
(being $Q^2$ and $A$ negative), and use Eq. (\ref{57}) to write the
potential proportionally to some power of the scalar field
\begin{equation}
V(Q)=-i^{-4/m}\left\{ 2^{-3-4/m}m^{2+4/m}(-\alpha A)^{-1-2/m}[-\alpha
A-6n(m-n)]{\alpha}^{2/m} \right\}Q^{2+4/m}\,.  \label{A3}
\end{equation}
Of course, we want $V$ to be a real function of $Q$, so that $4/m$ has
to be an integer. If it is an even number, so is also the exponent of
$Q$, and $i^{-4/m}Q^{2+4/m}$ is the product of two real quantities.
When $4/m$ is an odd number, such a product is still real, being both
the factors purely imaginary. We here choose to deal only with even
powers of $Q$, so neglecting the second possible situation and taking
$4/m$ even from now on.

Being $A<0$, it results $n<m(m+1)/(2-m)$ (including $m=2$), which soon
leads to values of the exponent $p\equiv2+4/m\neq4$. Moreover, due to
the condition $m\geq\tilde{m}$ (where $\sqrt{3}-1<\tilde{m}<2$), we
find $p>4$. This means that we have only one possible even value for
$p$, namely, $p_1\equiv6$; this value, of course, leads to a positive
power law potential
\begin{equation}
V(Q)=\frac{{\alpha}^2(-6n^2+7n-2)}{2^7(2-n)^3}Q^6\,,  \label{A4}
\end{equation}
where $1<n<2$, and we have $\sigma=n/m\sim1$, as required.

\subsubsection{$A=0$ (i.e., $m=\tilde{m}$)}

We have already noted that $m=\tilde{m}$ leads to a complete
dependence of the scalar-field time behavior on the matter content
through the $M$ parameter, since it is ${\dot{Q}}^2=-Mt^{-3n}$. From
$A=0$ (i.e., $m=\tilde{m}$), we get
\begin{equation}
n=\frac{m(1+m)}{2-m}\,,  \label{A5}
\end{equation}
with $\sqrt{3}-1<m<2$ (see above, after Eq. (\ref{56}). We also have
$\sigma=(1+m)/(2-m)\sim1.36$ for $m=\sqrt{3}-1$; thus, $m$ must stay
very near to this value.

Let us derive the form of the potential, firstly finding the form of the
imaginary $Q(t)$ and then writing time as (suitably choosing the sign)
\begin{equation}
t(Q)=(\frac{3n-2}{2i\sqrt{M}})^{\frac{2}{2-3n}}Q^{\frac{2}{2-3n}}\,,
\label{A7}
\end{equation}
(with $i$ the imaginary unit) which is a real and positive quantity,
being $Q^2<0$. Thus, we have
\begin{equation}
F(Q)=-\frac{1}{2\alpha}(\frac{3n-2}{2i\sqrt{M}})^{\frac{2m}{3n-2}}
Q^{\frac{2m}{3n-2}}  \label{A71}
\end{equation}
and, substituting $n$ from Eq. (\ref{A5}), the potential in Eq. (\ref{57})
becomes
\begin{equation}
V(Q)=\frac{\left[ \frac{i(2+m)}{C(m)}\sqrt{M}\right]^{C(m)}} {\alpha(m-2)^2}
D(m,\alpha,M) Q^{C(m)}\,,  \label{A8}
\end{equation}
where we have defined
\begin{equation}
C(m)\equiv\frac{8-2m^2}{3m^2+5m-4}\,,  \label{A9}
\end{equation}
\begin{equation}
D(m,\alpha,M)\equiv[3m^3+6m^4-4M\alpha +4mM\alpha-m^2(3+M\alpha)]\,.
\label{A9bis}
\end{equation}
Again, choosing only even integer exponents $C(m)$ for $Q$ gives real
values of the potential in Eq. (\ref{A8}). This gives the rather
strange values $m=\{(3\sqrt{17}-5)/8, (\sqrt{109}-5)/7,
(\sqrt{865}-15)/20 \}$. Moreover, we find that it is impossible to get
{\em both} $V$ and $F$ real. We conclude that this case is not only
little appealing, but also rather uninteresting from the physical
point of view.

\subsubsection{$m=3n-2$ (i.e., same powers of time)}

When $m=3n-2$, it can be shown that taking $n>1$ always leads to
$A<0$. Eliminating $m$ from the expression of $A$, we can get the
expression of $Q(t)$ and find that time is again a real and positive
function of $Q$
\begin{equation}
t(Q)= \left[ -i \frac{m}{2}{\alpha}^{1/2}(2-13n+12n^2+M\alpha)^{-1/2}
\right]^{-2/m} Q^{-2/m}\,.  \label{A11}
\end{equation}

The coupling function is
\begin{equation}
F(Q)=-\frac{1}{2}\left[ -\frac{m}{2}{\alpha}^{(m+2)/6m} \left( \frac{
4m^2+3m+3M\alpha-4}{3} \right)^{-1/2} \right]^2 Q^2\,,  \label{A111}
\end{equation}
and Eq. (\ref{57}) gives the real potential
\begin{equation}
V(Q)=-\left\{ \frac{m+6M\alpha-4}{6}\left[ -\frac{m}{2}{\alpha}
^{1/(m+2)}\left( \frac{4m^2+3m+3M\alpha-4}{3} \right)^{-1/2} \right]^{\frac{
2(m+2)}{m}} \right\}Q^{\frac{2(m+2)}{m}}\,,  \label{A12}
\end{equation}
where $m=3n-2$ is left as it is in order to have manageable
expressions. Although not dictated by the reality condition, we can
anyway ask for the exponent being an even integer $2K$. This gives
$m=2/(K-1)$ and $n=2K/[3(K-1)]$, so that $K\neq1$, $K<3$ (for $n>1$).
Moreover, we have $\sigma=K/3$, which limits $K$ to stay in the
interval $(1,3)$. The allowed values for $n$ are thus only $3$ and
$4/3$. While the first gives clearly nothing else but the case already
treated in Sec. III, the second seems to be untenable on observational
basis. Also, we have to be careful about the fact that this asymptotic
behavior in the neighbourhood of $\infty$ is not necessarily the same
as nowadays.

\subsection{Further general considerations}

There is something left to say in general in the case we are facing in this
section. Let us go back to our first general considerations and, starting
again from Eq. (\ref{4}) and taking Eq. (\ref{54}) into account, we can get
the function $p_{Q}=p_{Q}(t)$ in its full generality
\begin{equation}
p_{Q}(t) =-\frac{n(3n-2)}{\alpha}t^{-m-2}<0\,.  \label{64}
\end{equation}
We can soon see that such a result directly generalizes the one we
already found when $n=4/3$ and $m=2$. Also note that, again, we always
find $p_{Q}<0$. As to the dominant energy condition for $Q$, it is
noteworthy that it holds also in this more general situation, since we
always find
\begin{equation}
\rho_Q+p_{Q}=\left( \frac{2n}{\alpha}t^{3n-m-2}-M \right)t^{-3n}\geq0\,,
\label{64bis}
\end{equation}
due to previous results on physically allowed times. As before, we are
assuming $m\geq (m+1)/3>1$, so that $m>1$, too. Of course, we could also get
a DEC violation, once we accept $t^{3n-m-2}<M\alpha/(2n)$. But, as a matter
of fact, due to Eq. (\ref{51}) and being $M\alpha/(2n)<M\alpha/(3n^2)$, such
a violation could result only {\em together} with the violation of the
condition $\rho_Q>0$ on the $Q$-energy density. If we do not want to deal
with such a negative energy density, therefore, we have to conclude that the
situation we find here forbids a phantom cosmological behavior.

Let us compute, then, what happens to the scalar field equation of state. We
get
\begin{equation}
w_{Q}\equiv \frac{p_{Q}}{\rho_{Q}}=\frac{n(2-3n) t^{-m-2}}{
3n^{2}t^{-m-2}-\alpha Mt^{-3n}}\,,  \label{65}
\end{equation}
which is negative only at sufficiently late times, due again to Eq.
(\ref{54}). Note also that, differently from the special case treated
in Sec. III, this expression of the equation of state is now in
general non constant and, as $t\rightarrow \infty$, it is such that
$w_{Q}\rightarrow -1+2(3n)^{-1}$; since $n>1$, this means that the
greater is $n$ the more such an asymptotic value tends to $-1$, always
staying, however, $>-1$ and $<0$. On the other hand, Eq. (\ref{65})
yields
\begin{equation}
w_{Q}=-1\Longleftrightarrow t^{3n-m-2}=\frac{\alpha M}{2n}\,,  \label{66}
\end{equation}
that is, $w_{Q}=-1$ only at the well defined time
\begin{equation}
t^{*}\equiv \left( \frac{\alpha M}{2n}\right)^{\frac{1}{3n-m-2}}\,,
\label{67}
\end{equation}
such that $t^{*}>\bar{t}$ (the time from which $\rho_{Q}>0)$, as it could
already be expected. Also, it is $-1<w_{Q}<0$ when $t>t^{*}$, and then we
recover an {\em ordinary} scalar field behavior in an {\em ordinary}
cosmological context with dark energy after the decoupling epoch. Of course,
however, a family of $w<-1$ models is still viable, since we can have a DEC
violation when $\bar{t}<t<t^{*}$, that is, for a limited period of time,
during which it is anyway $\rho_Q<0$.

Finally, the matter density parameter becomes
\begin{equation}
\Omega_{m}\equiv \frac{G\rho_{m}}{3H^{2}}=\frac{\alpha M}{3n^{2}}
t^{m+2-3n}<1\,,  \label{68}
\end{equation}
implying that $\Omega_{m}=1/2$ only for $n=4/3$. It is interesting to note
here that $\Omega_{m}$ is now a decreasing function of time.

As a last remark, consider the expression of $\Lambda_{eff}$ due to
Eq. (\ref{5})
\begin{equation}
\Lambda_{eff}=\alpha \left( \frac{3n^{2}}{\alpha}t^{3n-m-2}-M\right) t^{m-3n}
\label{69}
\end{equation}
and compare it to the expected dependence $\Lambda_{eff}\sim H^{2}$,
which is in fact given by the situation described in the previous
section, as seen. Now, since $H^{2}\sim t^{-2}$ this means we have to
compare $t^{-2}$ and $t^{m-3n}$. As a matter of fact, taking Eq.
(\ref{54}) gives $m-3n\leq-2$, so that we can soon assert that we are
actually facing a sort of generalization of what is generally expected
and which we refer to in the Introduction. Also, note that we should
probably choose values of the parameters such that $3n\cong -m-2$. But
this would again imply constant values of $w_{Q}$ and $\Omega_{m}$,
being $p_{Q}\sim\rho_{Q}\sim \rho_{m}$.

Let us finally note that, due to Eq. (\ref{45}), it is $t\sim a^{1/n}$ and
\begin{equation}
\Lambda_{eff}(a) =\alpha \left( \frac{3n^{2}}{\alpha}a^{\frac{3n-m-2}{n}}
-M\right) a^{\frac{m-3n}{n}}\,,  \label{70}
\end{equation}
so that
\begin{equation}
\frac{d\ln \Lambda_{eff}}{d\ln a}=-\frac{2}{n}  \label{71}
\end{equation}
when $3n=m+2$. (This means that for $n=4/3$ and $m=2$, for example, we
find a decaying behavior of the cosmological constant as clearly
described in Eq. (\ref{70}) with $n=4/3$, i.e.,
$\Lambda_{eff}(a)=\Lambda_{0}a^{-3/2}$, $\Lambda_{0}$ being a
constant.)

\section{Conformal transformation}

In this section we make a sort of final comment. Our purpose is to
find the particular conformal transformation that connects the
nonminimally coupled model discussed in Sec. III with the usual
minimally coupled one we could write down in presence of both the
scalar field and dust. As a matter of fact, it is well known that such
a connection between, respectively, the {\em Jordan frame} and the
{\em Einstein frame} is indeed possible \cite{rug}. We here will limit
our considerations to the special transformation needed to go from one
framework to the other one.

Performing on the cosmological metric $g$ (in the Jordan frame) the
transformation
\begin{equation}
\hat{g}={f(Q)}^2 g  \label{72}
\end{equation}
($\hat{g}$ denoting corresponding quantities in the Einstein frame), with a
generic function $f$, our limited aim is then to determine explicitly the
form of this function. It is easy to see that $d\hat{t}=fdt$, $\hat{a}=fa$,
and the equations become \cite{rug}
\begin{equation}
3{\hat{H}}^{2}= {\hat{\rho}}_{m}+{\hat{\rho}}_{\hat{Q}}\,,  \label{73}
\end{equation}
\begin{equation}
\hat{Q}^{\prime\prime}+3\hat{H}\hat{Q}^{\prime}-\frac{d\hat{V}} {d\hat{Q}}-
\frac{df/d\hat{Q}}{\sqrt{1+6df/d\hat{Q}}}=0\,,  \label{74}
\end{equation}
where $8\pi G=1$ and prime denotes derivatives with respect to $\hat{t}$ and
\begin{equation}
{\hat{\rho}}_{m}= Mf^{-1}{\hat{a}}^{-3}\,,\,\,\,
d\hat{Q}=\sqrt{\frac{3(\frac{d F}{d Q})^2-F}{2F^2}}dQ\,,\,\,\,
\hat{V}=\frac{V}{4F^2}\,,  \label{75}
\end{equation}
being
\begin{equation}
f(Q)=\sqrt{-2F(Q)}\,.  \label{76}
\end{equation}
Note also that, as usual, in the Klein-Gordon equation for $Q(\hat{t})$ an
interaction term appears between the scalar field and the matter density
(through $f$).

If we use $Q(t)$ as defined in the Jordan frame (i.e., in Sec. III through
Eq. (\ref{25})), with $F=F(Q)$ as given in Eq. (\ref{28}), we soon find $f$
as a function of time
\begin{equation}
f(t)=\frac{t^{-1}}{\sqrt{\alpha}}\,,  \label{77}
\end{equation}
which in turn implies that the potential results constant
\begin{equation}
\hat{V}=\frac{\alpha (2-3M\alpha)}{6}\,.  \label{78}
\end{equation}
On the other hand, we can easily check the reality of the root term
for $d\hat{Q}$ in Eq. (\ref{75}), since it is
$[3(dF/dQ)^2-F]/(2F^2)=-M\alpha Q^{-2}>0$, being $Q$ an imaginary
quantity. We can also get the relation between the transformed time
$\hat{t}$ and $t$
\begin{equation}
\hat{t}=\frac{\ln t}{\sqrt{\alpha}}  \label{79}
\end{equation}
and find a de Sitter-like behavior of the scale factor in the Einstein frame
\begin{equation}
\hat{a}(\hat{t})=\frac{1}{\sqrt{\alpha}}\exp{(\frac{\sqrt{\alpha}\hat{t}}{3})}
\longrightarrow \hat{H}=\frac{\sqrt{\alpha}}{3}=const.\,;  \label{80}
\end{equation}
it is also
\begin{equation}
\hat{Q}=i \alpha\sqrt{M}\hat{t}\,,  \label{81}
\end{equation}
from which
\begin{equation}
\hat{Q}^{\prime} = i \alpha\sqrt{M}\hat{t}\,\,\,, \hat{Q}^{\prime\prime}
=0\,.  \label{82}
\end{equation}

As to the energy density of the scalar field in the Einstein frame, it also
results constant
\begin{equation}
\hat{\rho}_{\hat{Q}}\equiv \frac{1}{2}(\hat{Q}^{\prime})^2+\hat{V}
=\alpha\left( \frac{1}{3}-M\alpha \right)\,.  \label{83}
\end{equation}
Such a definition of the $\hat{Q}$-energy density is, of course,
different from the one we have used in the Jordan frame, since there
is no coupling now between geometry and $\hat{Q}$. The result shown in
Eq. (\ref{83}) soon gives a positive energy density for the
$\hat{Q}$-field when $M<(3\alpha)^{-1}$, which is allowed by the
constraint in Eq. (\ref{20}) for the positivity of the $Q$ energy
density. Note that, for $1/(3\alpha)<M<16/(3\alpha)$, negative values
of $\hat{\rho}_{\hat{Q}}$ can also be given.

Let us finally note that, at a first glance, things may appear simpler in
the Einstein frame, since there we get a constant potential for the scalar
field. However, there also appears a very uncomfortable interaction term
between $\hat{Q}$ and $M$ in the energy density term ${\hat{\rho}}_{m}$
(together with an additive term in the Klein-Gordon equation for the scalar
field). There is no apparent physical justification for such a situation.
This fact confirms, in our opinion, the nonequivalent settings of the theory
in the two different frameworks.

\section{Conclusive remarks}

In the sections above we have examined the consequences of some
assumptions on a standard nonminimally coupled cosmology, once we get
some inspiration for such assumptions from the modifications
introduced by the Renormalization Group (RG) equations in the Quantum
Einstein Gravity, with reference to works by M. Reuter and
collaborators \cite{3,4,5,9bis,agc}. This means that we have assumed
some power law behaviors for the scale factor and effective $G$ and
$\Lambda$ (according to the results of those papers), and we have
investigated their use inside our different situation. Both the
special case with a $4/3$-exponent and a generalized one have been
treated, showing that there is the possibility to get a phantom
cosmology, with an equation of state $w_Q<-1$. Among other things, it
seems interesting to note that this is not only connected to negative
kinetic energy terms for the scalar field, but also that, in some
situations, weak energy condition (WEC) violations become possible.

Most of all, we want to underline that the phantom behavior is
generally recovered for some ranges of values of the parameters used
in the models. This is already known as common in cosmology with
nonminimal coupling \cite{814}, but it is usually derived neglecting
the presence of other ingredients in the cosmic fluid besides the
scalar field. Here, we have instead found it with {\em both} the
scalar {\em and} the matter fields, also showing that the parameter
values allowing a phantom behavior are strictly constrained by the
presence of the parameter $M$, i.e., the present-day matter content.
Due to the still wide range of indeterminacy in the measures of the
dimensionless matter density parameter $\Omega_m$ \cite{schin}, this
in fact leads to a degeneration.

A comparison of Secs. III and IV shows that the case directly imported
from the RG improved setting is by far simpler and more elegant. The
exponent $4/3$ has also been tested against SNIa results (in fact the
older release)\cite{9bis}. We want to mention here that, in the
context of the usual minimal coupling setting, it is in fact possible
to find a general exact solution, which again shows an asymptotic
$t^{4/3}$ behavior \cite{grg}. This solution has been tested with a
full set of data \cite{grg,apj,prd}.

In other words, it appears like there is a sort of {\em conspiracy} in
favour of this value. The meaning of this, if any, is clearly still obscure
and deserves further investigation.

Also, some comments are in order about the work done in Ref.
\cite{reuterw}. It deals with the RG-improved gravity by replacing
$G_N$ and $\Lambda$ with scalar functions in the Einstein-Hilbert
action. The comparison with our work can be made posing
$g_{*}\lambda_{*}=-5/(2\pi M)$ and $\alpha=-6/M$, so implying
$\dot{Q}^2=0$, as can be easily seen from our Eq. (\ref{21}). This, of
course, leads to an impossibility of reconstructing the functions
$F=F(Q)$ and $V=V(Q)$. This suggests that the comparison is not
trivial and deserves further investigation.

Another aspect that, in our opinion, has to be emphasized is the always
decreasing behavior of $\rho_Q (t)$ in the Jordan frame, versus the constant
$\hat{\rho}_{\hat{Q}}$ in the Einstein frame, both in full contradiction
with what is commonly expected in minimally coupled cosmologies (where
usually a big rip is expected). It seems to us that one should probably be
cautious in studying approximated situations in the phase around the
present-day evolutionary stage of the universe, simply importing results and
expectations from minimally coupled cosmologies into those like the ones we
have treated here and, probably, in all kinds of scalar-tensor cosmologies.

On the other hand, let us stress again that the models treated in this paper
include only dust (ordinary nonrelativistic matter plus cold dark matter)
and scalar-field components, so that they can suitably describe the
cosmological situation only starting from the decoupling period. In this
epoch, the radiation component decreases much more than the matter one, and
has been completely neglected in our treatment. But what does it change in
the considerations above when we simply {\em replace} dust with radiation?
The situation so described is probably interesting only when the scalar
field can be better interpreted as an inflaton. (It could also be
investigated soon after inflation ends, when the scalar field could resemble
what remains of the inflaton {\em plus} what eventually can be guessed as
added to it in order to give the dark energy field much later emerging and
dominating the cosmic fluid content.) The problem is anyway analytically
treatable, at least for a situation similar to that dealt with in Sec. III,
but we postpone it to another forthcoming work. We can here simply note that
the coupling function $F(Q)$ now results always positive, so indicating that
$G_{eff}$ gives a repulsive force, while the potential $V(Q)$ is always a
negative quartic function of $Q$. We are in a regime where both the kinetic
energy and the pressure of the scalar field are always negative, but the
dominant energy condition can still be recovered. Finally, let us note that
we can get a sort of phantom cosmological behavior also in this situation.

\begin{table}[htbp]
\centering
\par
\begin{tabular}{|c|c|c|c|c|c|}
\hline
$0<M<\frac{2}{3\alpha }$ & $-1<w_{Q}<0$ & $\stackrel{{\cdot}}{Q}^{2}<0$ & $
\rho_{Q}>0$ & $\rho_{Q}+p_{Q}>0$ & $V>0$ \\ \hline
$\frac{2}{3\alpha }<M<\frac{8}{3\alpha }$ & $-1<w_{Q}<0$ &
$\stackrel{{\cdot} }{Q}^{2}<0$ & $\rho_{Q}>0$ & $\rho_{Q}+p_{Q}>0$ & $V<0$
\\ \hline $M=\frac{8}{3\alpha }$ & $w_{Q}=-1$ &
$\stackrel{{\cdot}}{Q}^{2}<0$ & $
\rho_{Q}>0$ & $\rho_{Q}+p_{Q}=0$ & $V<0$ \\ \hline
$\frac{8}{3\alpha }<M<\frac{16}{3\alpha }$ & $w_{Q}<-1$ &
$\stackrel{{\cdot}} {Q}^{2}<0$ & $\rho_{Q}>0$ & $\rho_{Q}+p_{Q}<0$ & $V<0$
\\ \hline
\end{tabular}
\par
\bigskip
\caption{Summary of the possible situations recovered in Sec. III. They
differ from one another depending on suitable ranges of the $M$ parameter,
characterizing the amount of ordinary matter in the universe.}
\end{table}

\end{document}